# Total Cross Section at Cosmic Ray Energies

Sohail Afzal Tahir[a], M. Ayub Faridi[a], Qadeer Afzal[a], Haris Rashid[a] and Fazal-e-Aleem[b]
*(a) Centre for High Energy Physics Punjab University Lahore, 54590 Pakistan.*
Presenter: M. Ayub Faridi  (ayubfaridi@yahoo.com), pak-faridi-M-abs1-he21-oral

Recent analysis of the cosmic ray data together with earlier experimental measurements at ISR and SPS provides us a sound footing to discuss the behavior of total cross section at asymptotic energies. We will study the growth of total cross section at high energies in the light of various theoretical approaches with special reference to measurements at RHIC and LHC.

## Introduction

Total cross section, $\sigma_T$ is one of the most fundamental parameters. In the hadronic scattering processes it has been studied extensively both in theory and experiment during the last three decades. In order to develop a theory, information about the rise of total cross section at cosmic energies would be very important [1-5]. Experimental information on the behavior of hadronic total cross sections at ultrahigh energies can be obtained from cosmic ray experiments.

Recent measurements at RHIC and LHC will take us in to the realm of cosmic ray energies where experimental information on *pp* total cross sections also exists from non-accelerator physics. However this information has large error bars in the numerical results. The information is therefore not conclusive in predicting the trend about the growth of the total cross section at cosmic ray energies. Consequently almost all the theoretical predictions fit this data. However, if we take into account the accurate data from RHIC, we can get useful information about the trend of total cross section at cosmic ray energies.

## Theoretical studies

A large amount of theoretical work has been undertaken by many authors which has been detailed in Refs. 1-12.  The work undertaken by these authors takes into account the trend of possible rise in total cross section and rho with different physical pictures which has been nicely worked out by Velasco et al [11] as shown in Figure 1.

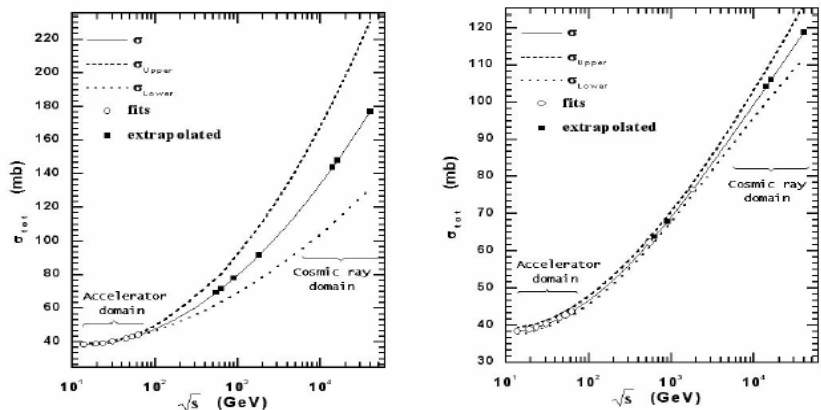

**Figure 1**

Keiji and Muneyuki in a recent work [7-8] have studied the predictions of proton(antiproton) proton cross section and rho rahio at LHC and cosmic-ray energy region. Total cross section at Tevatron-collider, LHC and cosmic-ray energies versus center of mass energy shown in Figure 2.

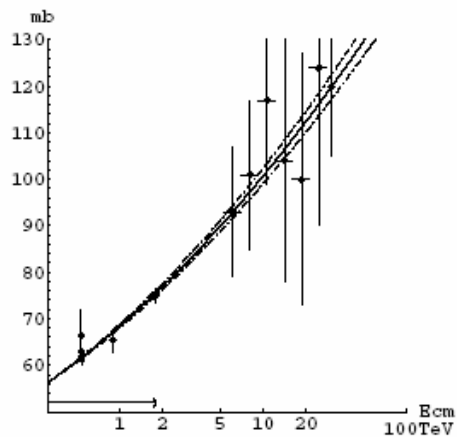

**Figure 2**

Fazal-e-Aleem and Sohail [1-5] in their recent work have taken up a detailed study of various aspects of proton (antiproton) proton scattering. It is observed that in almost all the models forward scattering amplitude parameters, $\sigma_T$ and $\rho$ are fitted well at the ISR energies. As we move to SPS Tevatron and cosmic ray energies, there is a difference of the predicted values. This difference becomes visible at LHC/Cosimc Ray energies. In *Regge models*, increase in the total cross section is approximated by the intercept of the Pomeron trajectory. High energy data is fitted well by this approximation although at ISR contributions from masonic trajectories is needed. The predicted cross section at 1.8 and 14 TeV is consistent with lns behaviour. The $\sigma_T$ value is predicted to be significantly higher when Odderon is taken in to account in the Regge framework. Predictions of the models with and without Odderon contribution differ in the RIHC and LHC region. In the *geometrical picture* total cross section is described by the shape of the colliding hadrons, which varies with energy. The geometrical picture gives a good fit to the experimental data for $\sqrt{s} > 20$ GeV. Real part of the radius (which has been taken as energy dependent) increases linearly with lns, which makes predictions to higher energy straightforward. Measurements of LHC will therefore give us a good indication of the trend for the total cross section. However, measurements in the near forward direction would be of significant importance at LHC as it would unambiguously establish or definitely contradict $(log\ s)^2$ behaviour which emerges as a consequence of Odderon. Various theoretical studies thus tell us that at LHC and Cosmic Ray energies predictions of different approaches are significantly different. A comparison of these studies [1-5] reveals that total cross sections values will begin to differ from the RHIC energies. More important would be the difference in the total cross section values for proton-proton and proton-antiproton scattering. This difference will become very prominent at the LHC and Cosmic Ray energies in case of the Odderon contribution. We also observe that the value of totoal cross section for different models varies from about 95 to about 145 mb at LHC. Although cosmic ray data due to large error bars accommodate these values, accurate measurements at LHC will be very important. We thus observe that what actually is the energy dependence of the total cross section is as yet without a definite answer. Currently data at ISR, SPS and Tevatron can be fitted with both lns and $ln^2s$ behaviour. On the one hand, rise of the total cross section is attributed to the exchange of soft Pomeron although the picture of Pomeron itself is not fully clear. On the other hand, rise of $\sigma_T$ is related to the shape of the colliding particle in the geometrical picture. In QCD inspired models the origin of this quantity is accounted for by the increase in the number of gluons. We also learn about some new and interesting interpretation of the Pomeron in the geometric picture of the diffractive scattering.